\documentclass[conference]{IEEEtran}

\usepackage{cite}
\usepackage{amsmath,amsfonts,amssymb}
\usepackage{graphicx}
\usepackage{array}
\usepackage{url}
\usepackage{amsthm}

\usepackage{subfig}
\usepackage{siunitx}
\usepackage{xcolor}

\newcommand{\hb}{\bar{\mathbf h}}
\newcommand{\Hb}{\hat{\bar{\mathbf h}}}
\newcommand{\E}{\mathbb E}
\newcommand{\CN}{\mathcal{CN}}
\newcommand{\diag}{\operatorname{diag}}

\begin{document}

\title{Spectral Efficiency Analysis of Massive MIMO-OFDM Systems with Double Quantization}

\author{\IEEEauthorblockN{Bengu Bilgic Keskin\textsuperscript{1}, A. Oguz Kislal\textsuperscript{1}, Ali Gorcin\textsuperscript{1,2},  Özlem Tuğfe Demir\textsuperscript{3}}
\IEEEauthorblockA{\textsuperscript{1}\textit{Communications and Signal Processing Research (HISAR) Lab, TUBITAK BILGEM, Kocaeli, Turkiye}\\
\textsuperscript{2}\textit{Faculty of Electrical and Electronics Engineering, Istanbul Technical University, Istanbul, Turkiye} \\
\textsuperscript{3}\textit{Department of Electrical and Electronics Engineering, Bilkent University, Ankara, Turkiye} \\ 
Emails: \{bengu.bilgic, ahmet.kislal\}@tubitak.gov.tr, aligorcin@itu.edu.tr, ozlemtugfedemir@bilkent.edu.tr}
}

\maketitle

\begin{abstract}
Massive multiple-input multiple-output (MIMO) achieves a high spectral efficiency (SE) by employing a large number of receive antennas in the uplink. Employing many receive antennas requires a large number of analog-to-digital converters (ADCs) and a high fronthaul data rate. Low-resolution ADCs can reduce power consumption and cost, while fronthaul quantization can reduce the required fronthaul data rate, at the expense of quantization distortion at both stages.
In wideband orthogonal frequency-division multiplexing systems, ADC quantization is performed in the time domain, while fronthaul quantization may be applied after the discrete Fourier transform to transmit only active subcarriers. In this study, we investigate the achievable uplink SE over fronthaul links under double quantization and imperfect channel estimation. We derive a linear minimum mean-squared error channel estimator from the double-quantized pilot observations and obtain an achievable SE using the use-and-then-forget bound. Numerical results demonstrate that the ADC and fronthaul quantization resolutions have comparable impacts on the achievable SE. When the two resolutions differ, the lower resolution becomes the dominant factor limiting the SE. Furthermore, our numerical results show that six-bit ADC and fronthaul quantization achieve more than 90\% of the SE attained in the ideal case.
\end{abstract}

\begin{IEEEkeywords}
Massive MIMO, wideband OFDM, low-resolution ADC, fronthaul quantization, hardware impairments
\end{IEEEkeywords}

\vspace{-2mm}
\section{Introduction}
\label{sec:intro}

  \par Massive multiple-input multiple-output (MIMO) is a key physical-layer technology in next-generation cellular systems \cite{massive_mimo_foundations}. By equipping a base station (BS) with a large number of antennas, massive MIMO can simultaneously serve multiple user equipments (UEs) over the same time and frequency resources and achieve high spectral efficiency \cite{massive_mimo_foundations}. As wireless networks evolve toward 6G, massive MIMO is expected to remain a key technology for addressing growing traffic demands and increasingly stringent performance requirements \cite{towards_6g_mimo}.

\par The practical implementation of massive MIMO becomes challenging as the number of BS antennas increases. In particular, using many high-resolution analog-to-digital converters (ADCs) has substantial power consumption and cost \cite{aqnm_rician_se}. Low-resolution ADCs are attractive in this sense, but they introduce quantization distortion \cite{aqnm_massive_mimo_rate}.

In massive MIMO-OFDM systems, ADC quantization is performed on the time-domain samples before the discrete Fourier transform (DFT) \cite{wideband_quantized_mimo_ofdm}. To accommodate the fronthaul data rates demanded by massive MIMO systems, the null subcarriers are discarded after the DFT, and only the samples on the active subcarriers are transmitted over the fronthaul link. To further reduce the required fronthaul data rate, these samples are quantized before transmission. However, this comes with a trade-off as an additional distortion is introduced to the system depending on the fronthaul quantization resolution \cite{fronthaul_quantization}.


\par Low-resolution ADCs have been considered in several studies on massive MIMO-OFDM systems. Channel estimation and data detection algorithms are developed in \cite{wideband_quantized_mimo_ofdm}, an achievable sum-rate expression is derived for the oversampled receiver in \cite{oversampled_quantized_mimo_ofdm}, and cell-free massive MIMO-OFDM is considered in \cite{wideband_cell_free_adc}. However, these works do not consider non-ideal fronthaul quantization.

\par Fronthaul constraints have also been considered in massive MIMO systems. The effect of a fronthaul rate constraint when low-resolution data converters are employed is studied in \cite{fronthaul_rate_constraint_quantized_mimo}, fronthaul quantization in cell-free massive MIMO is considered in \cite{cell_free_uniform_quantization}, low-resolution data converters and fronthaul compression are considered in \cite{cell_free_mmwave_adc_fronthaul}, and fronthaul quantization with low-resolution ADCs is studied in \cite{fronthaul_quantization}. However, these works do not consider wideband massive MIMO-OFDM systems in which both time-domain ADC and frequency-domain fronthaul quantization are present, giving rise to what we refer to as the \emph{double quantization} effect.
In \cite{double_quantization_ofdm}, a wideband cell-free system with ADC and fronthaul quantization is considered, and a quantization-aware signal reconstruction strategy is introduced. However, this work does not provide an information-theoretic analysis of this system.

The main contributions of this paper are summarized as follows:
\begin{itemize}

\item We develop an analytical framework for wideband massive MIMO-OFDM systems subject to \emph{double quantization}, where the received time-domain samples are first quantized by low-resolution ADCs and, after the DFT and removal of inactive subcarriers, the active frequency-domain samples are quantized again for fronthaul transmission.
The framework explicitly captures the distortion introduced by both
quantization stages.

\item We derive a linear minimum mean-squared error (LMMSE) channel estimator based on double-quantized
pilot observations. The resulting estimator accounts for the effective
noise generated jointly by thermal noise, time-domain ADC quantization,
and frequency-domain fronthaul quantization, and accommodates pilot
contamination caused by pilot reuse.

\item We derive an achievable uplink SE expression under imperfect
channel estimation using the use-and-then-forget (UatF) bound.
The proposed analysis incorporates both quantization stages into the
effective interference-and-noise statistics and is applicable to
general linear combining schemes.

\item Through numerical evaluations with normalized maximum ratio (NMR), regularized zero-forcing (RZF), and minimum mean-squared error (MMSE) combining,
we characterize the joint impact of ADC and fronthaul quantization
resolutions. The results reveal a bottleneck effect: increasing the
resolution of one quantization stage provides only limited SE gains
when the other stage operates at low resolution. Consequently, balanced
ADC and fronthaul resolutions are generally more effective than
over-provisioning either stage alone. In particular, using six-bit
quantization at both stages achieve more than 90\% of the SE attained in the unquantized benchmark in the considered setup.

\end{itemize}

\vspace{-2mm}
\section{System Model}
\label{sec:model}

\par We consider the uplink of a massive MIMO-OFDM system in which $K$ single-antenna UEs communicate with a BS equipped with $N$ receive antennas. 
The time-domain signal is first quantized by low-resolution ADCs and transformed to the frequency domain. The inactive subcarriers are then discarded, and the samples on the active subcarriers are quantized for transmission over the
fronthaul link. Channel estimation and data detection are performed on the fronthaul output. Both quantization stages are characterized by the additive quantization noise model (AQNM).
\vspace{-2mm}
\subsection{Signal and Channel Model}
 
\par The system employs OFDM with $M$ subcarriers. Without loss of generality, we consider an arbitrary OFDM symbol allocated for uplink data transmission. UE $i$ transmits with power $\rho_i$ over an $R$-tap frequency-selective channel. The time-domain OFDM sample of UE $i$ is given by
\begin{equation}
s_i[q]
=
\frac{1}{\sqrt M}
\sum_{m=0}^{M-1}
\bar s_i[m]e^{j2\pi qm/M},
\end{equation}
where $q=-(R-1),\dots,M-1$, and a cyclic prefix (CP) of length $R-1$ is appended to the time-domain samples. Here, $\bar s_i[m]$ is the data symbol of UE $i$ on subcarrier $m$, and the data symbols are independent across UEs and subcarriers. After CP removal, the unquantized signal received at the BS antennas is
\begin{equation}
\mathbf y[q]
=
\sum_{i=1}^{K}
\sum_{r=0}^{R-1}
\mathbf h_i[r]\sqrt{\rho_i}\,s_i[q-r]
+
\mathbf w[q],
\label{eq:y_time_unquantized}
\end{equation}
where $q=0,\dots,M-1$, $\mathbf h_i[r]\in\mathbb C^{N\times1}$ is the $r$-th channel tap between UE $i$ and the BS, and
$\mathbf w[q]\sim\CN(\mathbf 0,\sigma^2\mathbf I_N)$ is the additive noise.

\par The channel taps follow a spatially correlated Rayleigh fading model:
\begin{equation}
\mathbf h_i[r]\sim\CN(\mathbf 0,\mathbf R_{i,r}),
\qquad r=0,\dots,R-1,
\end{equation}
where $\mathbf R_{i,r}$ is the spatial correlation matrix of the $r$th tap and $h_i[r]$ is mutually independent with respect to both $r$ and $i$. The corresponding frequency-domain channel is obtained as
\begin{equation}
\hb_i[m]
=
\sum_{r=0}^{R-1}
\mathbf h_i[r]e^{-j2\pi rm/M}.
\end{equation}
It follows that $\hb_i[m]\sim\CN(\mathbf 0,\mathbf R_i)$ with
$\mathbf R_i=\sum_{r=0}^{R-1}\mathbf R_{i,r}$.
\par To maintain analytical tractability while preserving the frequency-selective nature of the underlying channel, we adopt a coherence-block (CB) approximation in the frequency domain \cite{coherence_block_ofdm}. The $M$ subcarriers are
divided into $L_{\mathrm{CB}}$ CBs, each containing $\lambda_{\mathrm{CB}}$
consecutive subcarriers, so that $L_{\mathrm{CB}}=~M/\lambda_{\mathrm{CB}}$.
Let $\mathcal M_l$ denote the subcarrier set of the $l$th CB. Within each CB, the frequency-domain channel is represented as constant, i.e.,
$\hb_i[m]=\hb_{i,l}$ for $m\in\mathcal M_l$, where
$\hb_{i,l}\sim\CN(\mathbf 0,\mathbf R_i)$.  Each CB spans $\nu_{\mathrm{CB}}$ consecutive OFDM symbols in time and the $\lambda_{\mathrm{CB}}$ subcarriers in $\mathcal M_l$ in frequency. The representative channel vectors corresponding to different CBs are modeled as statistically independent.
\par Only a subset of the CBs is allocated to the uplink transmission. Let
$\mathcal A\subseteq\{1,\dots,L_{\mathrm{CB}}\}$ denote the set of active CBs,
with $L_{\mathrm{act}}\triangleq|\mathcal A|$, and let
\begin{equation}
\beta
\triangleq
\frac{L_{\mathrm{act}}}{L_{\mathrm{CB}}}
=
\frac{L_{\mathrm{act}}\lambda_{\mathrm{CB}}}{M}
\label{eq:beta_def}
\end{equation}
denote the spectral occupancy ratio. The UEs transmit only on the subcarriers of the active CBs, i.e., $\bar s_i[m]=0$ for $m\in\mathcal M_l$ with $l\notin\mathcal A$, and $\E\{|\bar s_i[m]|^2\}=1$ for $m\in\mathcal M_l$ with $l\in\mathcal A$. Setting $\beta=1$ corresponds to the case in which all subcarriers are used.
\par Using the CB representation, the received signal in
\eqref{eq:y_time_unquantized} can be written as
\begin{equation}
\mathbf y[q]
=
\sum_{i=1}^{K}
\sum_{l\in\mathcal A}
\hb_{i,l}\sqrt{\rho_i}\,\xi_{i,l}[q]
+
\mathbf w[q],
\label{eq:y_cb}
\end{equation}
where
\begin{equation}
\xi_{i,l}[q]
\triangleq
\frac{1}{\sqrt M}
\sum_{m\in\mathcal M_l}
\bar s_i[m]e^{j2\pi qm/M}
\label{eq:xi_def}
\end{equation}
is the time-domain contribution of UE $i$ from the subcarriers in $\mathcal M_l$.

\subsection{ADC Quantization}
\par The received signal is quantized by the low-resolution ADCs at the BS. Using the AQNM \cite{bussgang_aqnm,aqnm_massive_mimo_rate}, the ADC output is given by
\begin{equation}
\mathbf y'[q]
=
\alpha_1\mathbf y[q]
+
\boldsymbol\eta_1[q],
\label{eq:adc_output}
\end{equation}
where $\alpha_1\in(0,1]$ is the ADC quantization gain and $\boldsymbol\eta_1[q]$ is the ADC quantization distortion. The distortion is uncorrelated with $\mathbf y[q]$ \cite{bussgang_aqnm} and is modeled as

\begin{equation}
\boldsymbol\eta_1[q]\mid\mathcal H
\sim
\CN\left(\mathbf 0,\mathbf D_{\eta_1\mid\mathcal H}\right),
\label{eq:eta1_distribution}
\end{equation}
where
$\mathcal H\triangleq\{\hb_{i,l}:i=1,\dots,K,\ l\in\mathcal A\}$
denotes the set of channel realizations over the active CBs. Its conditional
covariance is modeled as
\cite{fronthaul_quantization}
\begin{equation}
\mathbf D_{\eta_1\mid\mathcal H}
=
\alpha_1(1-\alpha_1)
\diag\!\left(
\mathbf C_{\mathbf y\mathbf y\mid\mathcal H}
\right),
\label{eq:Reta1_cond}
\end{equation}
where
$\mathbf C_{\mathbf y\mathbf y\mid\mathcal H}\triangleq\E\{\mathbf y[q]\mathbf y[q]^{\mathrm H}\mid\mathcal H\}$
is the conditional covariance of $\mathbf y[q]$, which is given as
\begin{equation}
\mathbf C_{\mathbf y\mathbf y\mid\mathcal H}
=
\sum_{i=1}^{K}
\sum_{l'\in\mathcal A}
\rho_i\frac{\lambda_{\mathrm{CB}}}{M}
\hb_{i,l'}\hb_{i,l'}^{\mathrm H}
+
\sigma^2\mathbf I_N.
\label{eq:Cyy_cond}
\end{equation}
Here, \eqref{eq:Cyy_cond} follows from
$\E\{\xi_{i,l}[q]\xi_{j,l'}[q]^*\}
=\delta_{ij}\delta_{ll'}\lambda_{\mathrm{CB}}/M$ and the independence between
the data symbols and the noise.

\par The BS applies $M$-point DFT to the ADC output in
\eqref{eq:adc_output}. For a subcarrier $m\in\mathcal M_l$ of an active CB
$l\in\mathcal A$, the frequency-domain received signal after ADC quantization
is
\begin{equation}
\bar{\mathbf y}'[m]
=
\alpha_1
\sum_{i=1}^{K}
\hb_{i,l}\sqrt{\rho_i}\,\bar s_i[m]
+
\alpha_1\bar{\mathbf w}[m]
+
\bar{\boldsymbol\eta}_1[m],
\label{eq:ybar_after_adc}
\end{equation}
where $\bar{\mathbf w}[m]\sim\CN(\mathbf 0,\sigma^2\mathbf I_N)$ is the additive noise in the frequency-domain and
\begin{equation}
\bar{\boldsymbol\eta}_1[m]
\triangleq
\frac{1}{\sqrt M}
\sum_{q=0}^{M-1}
\boldsymbol\eta_1[q]e^{-j2\pi qm/M}
\label{eq:eta1_dft}
\end{equation}
is the frequency-domain ADC distortion. The time-domain distortion samples are modeled as uncorrelated, i.e., $\E\{\boldsymbol\eta_1[q]\boldsymbol\eta_1[q']^{\mathrm H}\mid\mathcal H\}=\mathbf D_{\eta_1\mid\mathcal H}\,\delta_{qq'}$. Hence, the frequency-domain ADC distortion is uncorrelated across subcarriers and has the same conditional covariance as the time-domain distortion as
\begin{equation}
\mathbf D_{\bar\eta_1\mid\mathcal H}
\triangleq
\E\left\{
\bar{\boldsymbol\eta}_1[m]\bar{\boldsymbol\eta}_1[m]^{\mathrm H}
\mid\mathcal H
\right\}
=
\mathbf D_{\eta_1\mid\mathcal H}.
\label{eq:Rbareta1_cond}
\end{equation}

\subsection{Fronthaul Quantization}

\par The ADC-quantized frequency-domain samples corresponding to the active CBs are forwarded over the fronthaul link, where the second quantization stage is applied. Since the subcarriers of the inactive CBs carry only noise and distortion, they are discarded before the fronthaul quantization. As a result, the number of quantized samples per OFDM symbol is reduced to $\beta M$. Using the AQNM for the fronthaul quantizer \cite{bussgang_aqnm,fronthaul_quantization}, the fronthaul output is given by
\begin{equation}
\bar{\mathbf y}''[m]
=
\alpha_2\bar{\mathbf y}'[m]
+
\bar{\boldsymbol\eta}_2[m],
\label{eq:fronthaul_output}
\end{equation}
where $\alpha_2\in(0,1]$ is the fronthaul quantization gain and
$\bar{\boldsymbol\eta}_2[m]$ is the fronthaul quantization distortion. Substituting
\eqref{eq:ybar_after_adc} into \eqref{eq:fronthaul_output}, the double-quantized
received signal is
\begin{equation}
\bar{\mathbf y}''[m]
=
\alpha_2\alpha_1
\sum_{i=1}^{K}
\hb_{i,l}\sqrt{\rho_i}\,\bar s_i[m]
+
\tilde{\mathbf n}[m],
\label{eq:rx}
\end{equation}
where
\begin{equation}
\tilde{\mathbf n}[m]
\triangleq
\alpha_2\alpha_1\bar{\mathbf w}[m]
+
\alpha_2\bar{\boldsymbol\eta}_1[m]
+
\bar{\boldsymbol\eta}_2[m]
\label{eq:effective_noise}
\end{equation}
is the effective noise.

\par Similar to the ADC quantization model, the fronthaul distortion is
modeled as
\begin{equation}
\bar{\boldsymbol\eta}_2[m]\mid\mathcal H
\sim
\CN\left(\mathbf 0,\mathbf D_{\bar\eta_2\mid\mathcal H}\right),
\label{eq:eta2_distribution}
\end{equation}
which is uncorrelated with the fronthaul input $\bar{\mathbf y}'[m]$ and across subcarriers. We further model the distortion terms introduced by the two quantization stages and the additive noise as mutually uncorrelated. Consequently, the components of the
effective noise in \eqref{eq:effective_noise} are mutually uncorrelated. The covariance of the fronthaul distortion is\footnote{Note that fronthaul distortion covariance $\mathbf D_{\bar\eta_2\mid\mathcal H}$ depends on the CB via $\hb_{i,l}$. This dependence is omitted from the notation for brevity.}
\begin{equation}
\mathbf D_{\bar\eta_2\mid\mathcal H}
=
\alpha_2(1-\alpha_2)
\diag\!\left(
\mathbf C_{\bar{\mathbf y}'\bar{\mathbf y}'\mid\mathcal H}
\right),
\label{eq:Reta2_cond}
\end{equation}
where
$\mathbf C_{\bar{\mathbf y}'\bar{\mathbf y}'\mid\mathcal H}
\triangleq
\E\{\bar{\mathbf y}'[m]\bar{\mathbf y}'[m]^{\mathrm H}\mid\mathcal H\}$
is the conditional covariance of the fronthaul input for $m\in\mathcal M_l$ with
$l\in\mathcal A$. This covariance is obtained as
\begin{equation}
\mathbf C_{\bar{\mathbf y}'\bar{\mathbf y}'\mid\mathcal H}
=
\alpha_1^2
\sum_{i=1}^{K}
\rho_i\hb_{i,l}\hb_{i,l}^{\mathrm H}
+
\alpha_1^2\sigma^2\mathbf I_N
+
\mathbf D_{\bar\eta_1\mid\mathcal H}.
\label{eq:Cyyp_l_cond}
\end{equation}

\par The quantizer gains $\alpha_1$ and $\alpha_2$ are determined by the resolutions of the ADC and fronthaul quantizers, respectively. For a $b$-bit quantizer, the gain $\alpha$ takes the values in Table~\ref{tab:alpha} for
$b\le5$, and is computed as $\alpha=1-\frac{\pi\sqrt3}{2}\,2^{-2b}$ for higher
resolutions \cite{aqnm_massive_mimo_rate}. $\alpha=1$ corresponds to the unquantized case.
\begin{table}[t]
\centering
\vspace{0.13in}
\caption{Values of the quantizer gain $\alpha$ for $1$--$5$ bits.}
\label{tab:alpha}
\begin{tabular}{c|ccccc}
\hline
$b$ & 1-bit & 2-bit & 3-bit & 4-bit & 5-bit\\
\hline
$\alpha$ & 0.6366 & 0.8825 & 0.96546 & 0.990503 & 0.997501\\
\hline
\end{tabular}
\end{table}

\section{Channel Estimation}
\label{sec:est}

\subsection{Pilot Transmission}

\par In each active CB $l\in\mathcal A$, one OFDM symbol over the subcarriers in
$\mathcal M_l$ is allocated to pilot transmission. The UEs are assigned pilot sequences of length $\tau_p=\lambda_{\mathrm{CB}}$ from a set $\{\boldsymbol\phi_1,\dots,\boldsymbol\phi_{\tau_p}\}$ satisfying
$\boldsymbol\phi_n^{\mathrm H}\boldsymbol\phi_{n'}=\tau_p\delta_{n,n'}$.
Specifically, the sequence assigned to UE $i$ is $\boldsymbol\phi_{t_i}$, where
the pilot index $t_i\in\{1,\dots,\tau_p\}$ is given by
$t_i=(\bmod_{\tau_p} (i-1))+1$. The entries of $\boldsymbol\phi_{t_i}$ are the pilot symbols transmitted by UE
$i$, with $\phi_{t_i}[m]$ denoting the symbol correspoding to $m$th subcarrier. UE $i$ transmits its pilot symbols with power $\rho_i$.

\par The received pilot signal before ADC quantization is
\begin{equation}
\mathbf y_{\mathrm p}[q]
=
\sum_{i=1}^{K}
\sum_{l\in\mathcal A}
\hb_{i,l}\sqrt{\rho_i}\,\psi_{i,l}[q]
+
\mathbf w_{\mathrm p}[q],
\label{eq:y_pilot_time}
\end{equation}
where
\begin{equation}
\psi_{i,l}[q]
\triangleq
\frac{1}{\sqrt M}
\sum_{m\in\mathcal M_{l}}
\phi_{t_i}[m]e^{j2\pi qm/M}
\label{eq:psi_def}
\end{equation}
is the time-domain contribution of the pilot symbols of UE $i$ from the subcarriers in $\mathcal M_{l}$ and
$\mathbf w_{\mathrm p}[q]\sim\CN(\mathbf 0,\sigma^2\mathbf I_N)$. Similar to \eqref{eq:rx}, the double-quantized received pilot signal on subcarrier $m\in\mathcal M_l$ is
\begin{equation}
\bar{\mathbf y}_{\mathrm p}''[m]
=
\alpha_2\alpha_1
\sum_{i=1}^{K}
\hb_{i,l}\sqrt{\rho_i}\,\phi_{t_i}[m]
+
\tilde{\mathbf n}_{\mathrm p}[m],
\label{eq:pilot_rx_freq}
\end{equation}
where
\begin{equation}
\tilde{\mathbf n}_{\mathrm p}[m]
\triangleq
\alpha_2\alpha_1\bar{\mathbf w}_{\mathrm p}[m]
+
\alpha_2\bar{\boldsymbol\eta}_{1,\mathrm p}[m]
+
\bar{\boldsymbol\eta}_{2,\mathrm p}[m]
\label{eq:pilot_effective_noise_freq}
\end{equation}
is the effective noise during pilot transmission. Here,
$\bar{\mathbf w}_{\mathrm p}[m]\sim\CN(\mathbf 0,\sigma^2\mathbf I_N)$,
$\bar{\boldsymbol\eta}_{1,\mathrm p}[m]$ is the frequency-domain ADC
distortion, and $\bar{\boldsymbol\eta}_{2,\mathrm p}[m]$ is the fronthaul
distortion. These three terms are mutually uncorrelated, and
$\tilde{\mathbf n}_{\mathrm p}[m]$ is uncorrelated across subcarriers.

\par To evaluate the distortion covariance matrices during pilot transmission,
we use the Gaussian-equivalent pilot representation in \cite{massive_mimo_foundations}, where a Haar-distributed unitary rotation is applied to the deterministic pilot
sequences while preserving their orthogonality. This yields
$\E\{\psi_{t_i,l}[q]\psi_{t_j,l'}[q]^*\}
~=~\delta_{t_i t_j}\delta_{ll'}\lambda_{\mathrm{CB}}/M$ for $l,l'\in\mathcal A$,
and hence $\E\{\mathbf y_{\mathrm p}[q]\mathbf y_{\mathrm p}^{\mathrm H}[q]\mid\mathcal H\}
=\mathbf C_{\mathbf y\mathbf y\mid\mathcal H}$. Therefore, the AQNM distortion covariances during pilot and data transmission are identical, i.e., 
$\mathbf D_{\bar\eta_{1,\mathrm p}\mid\mathcal H}
=\mathbf D_{\bar\eta_1\mid\mathcal H}$ and
$\mathbf D_{\bar\eta_{2,\mathrm p}\mid\mathcal H}=\mathbf D_{\bar\eta_{2}\mid\mathcal H}.$

\subsection{LMMSE Channel Estimation}

\par The LMMSE estimate of $\hb_{k,l}$ is given by
\begin{equation}
\Hb_{k,l}
=
\E\left\{
\hb_{k,l}\mathbf y_{t_k,l}^{\mathrm H}
\right\}
{\E\left\{
\mathbf y_{t_k,l}\mathbf y_{t_k,l}^{\mathrm H}
\right\}}^{-1} \mathbf{y}_{t_k,l}
.
\label{eq:lmmse_general}
\end{equation}
Here,
$\mathbf y_{t_k,l}
=
\frac{1}{\sqrt{\tau_p}}
\sum_{m\in\mathcal M_l}
\bar{\mathbf y}_{\mathrm p}''[m]\phi_{t_k}^*[m]$
denotes the pilot observation obtained over the subcarriers in
$\mathcal M_l$. Using the orthogonality of the pilot sequences, this
observation can be rewritten as
\begin{equation}
\mathbf y_{t_k,l}
=
\alpha_2\alpha_1
\sum_{i\in\mathcal P_k}
\sqrt{\rho_i\tau_p}\,\hb_{i,l}
+
\tilde{\mathbf n}_{t_k,l},
\label{eq:pilot_observation}
\end{equation}
where
$\mathcal P_k\triangleq\{i:t_i=t_k\}$
is the set of UEs sharing the pilot of UE $k$, and
$\tilde{\mathbf n}_{t_k,l}
\triangleq
\frac{1}{\sqrt{\tau_p}}
\sum_{m\in\mathcal M_l}
\tilde{\mathbf n}_{\mathrm p}[m]\phi_{t_k}^*[m]$
is the effective noise in the pilot observation. The expectations required in
\eqref{eq:lmmse_general} can then be obtained from
\eqref{eq:pilot_observation} as
\begin{equation}
\E\left\{
\hb_{k,l}\mathbf y_{t_k,l}^{\mathrm H}
\right\}
=
\alpha_2\alpha_1
\sqrt{\rho_k\tau_p}\,\mathbf R_k
\label{eq:cross_cov_h_y}
\end{equation}
and
\begin{equation}
\begin{aligned}
\E\left\{
\mathbf y_{t_k,l}\mathbf y_{t_k,l}^{\mathrm H}
\right\}
&=
(\alpha_2\alpha_1)^2
\sum_{i\in\mathcal P_k}
\rho_i\tau_p\mathbf R_i
+
\mathbf C_{\tilde{\mathbf n}_{t_k}\tilde{\mathbf n}_{t_k}},
\label{eq:pilot_obs_cov}
\end{aligned}
\end{equation}
where
\begin{equation}
\begin{aligned}
\mathbf C_{\tilde{\mathbf n}_{t_k}\tilde{\mathbf n}_{t_k}}
&\triangleq
\E\left\{
\tilde{\mathbf n}_{t_k,l}
\tilde{\mathbf n}_{t_k,l}^{\mathrm H}
\right\} \\
&=
(\alpha_2\alpha_1)^2\sigma^2\mathbf I_N
+
\alpha_2^2\mathbf D_{\bar\eta_{1}}
+
\mathbf D_{\bar\eta_{2}}
\end{aligned}
\label{eq:Ctk}
\end{equation}
is the covariance of the effective noise in the pilot observation. Here,
$\mathbf D_{\bar\eta_{1}}
\triangleq
\E_{\mathcal H}
\left\{
\mathbf D_{\bar\eta_{1}\mid\mathcal H}
\right\}$
and
$\mathbf D_{\bar\eta_{2}}
\triangleq
\E_{\mathcal H}
\left\{
\mathbf D_{\bar\eta_{2}\mid\mathcal H}
\right\}$
are the average ADC and fronthaul distortion covariances, respectively. By taking the expectation over channel realizations in \eqref{eq:Reta1_cond} and \eqref{eq:Reta2_cond}, $\mathbf D_{\bar\eta_{1}}$ and $\mathbf D_{\bar\eta_{2}}$ can be found as
\begin{equation}
\mathbf D_{\bar\eta_{1}}
=
\alpha_1(1-\alpha_1)
\diag\!\left(
\beta
\sum_{i=1}^{K}
\rho_i\mathbf R_i
+
\sigma^2\mathbf I_N
\right),
\label{eq:Rbareta1_pilot}
\end{equation}
\begin{equation}
\mathbf D_{\bar\eta_{2}}
=
\alpha_2(1-\alpha_2)
\diag\!\left(
\alpha_1^2
\sum_{i=1}^{K}
\rho_i\mathbf R_i
+
\alpha_1^2\sigma^2\mathbf I_N
+
\mathbf D_{\bar\eta_{1}}
\right).
\label{eq:Rbareta2_pilot}
\end{equation}
By substituting \eqref{eq:cross_cov_h_y} and
\eqref{eq:pilot_obs_cov} into \eqref{eq:lmmse_general}, the LMMSE
estimate is given by
\begin{equation}
\Hb_{k,l}
=
\alpha_2\alpha_1
\sqrt{\rho_k\tau_p}\,
\mathbf R_k
\boldsymbol\Psi_k
\mathbf y_{t_k,l},
\label{eq:lmmse}
\end{equation}
where $\boldsymbol\Psi_k = \E\left\{
\mathbf y_{t_k,l}\mathbf y_{t_k,l}^{\mathrm H}
\right\}^{-1}$ and the channel estimation-error covariance is
\begin{equation}
\mathbf C_{\mathrm{err},k}
=
\mathbf R_k
-
(\alpha_2\alpha_1)^2
\rho_k\tau_p\,
\mathbf R_k
\boldsymbol\Psi_k
\mathbf R_k.
\label{eq:errcov}
\end{equation}
For $\alpha_1=\alpha_2=1$, the estimator in \eqref{eq:lmmse} reduces to the MMSE estimator of the unquantized system.

\section{Uplink Data Detection and Spectral Efficiency}
\label{sec:se}
 
\par For each subcarrier $m\in\mathcal M_l$ of an active CB $l\in\mathcal A$, a linear combining vector $\mathbf v_{k,l}\in\mathbb C^{N\times 1}$ is applied to the
double-quantized received signal in \eqref{eq:rx}. The resulting signal
for UE $k$ is
\begin{equation}
\hat{\bar s}_k[m]
=
\mathbf v_{k,l}^{\mathrm H}\bar{\mathbf y}''[m].
\label{eq:data_detection}
\end{equation}
Substituting \eqref{eq:rx} into \eqref{eq:data_detection} yields
\begin{equation}
\hat{\bar s}_k[m]
=
\alpha_2\alpha_1
\sum_{i=1}^{K}
\mathbf v_{k,l}^{\mathrm H}\hb_{i,l}
\sqrt{\rho_i}\,\bar s_i[m]
+
\mathbf v_{k,l}^{\mathrm H}\tilde{\mathbf n}[m].
\label{eq:combiner_output}
\end{equation}
Since the pilot observation in \eqref{eq:pilot_observation} is generally not Gaussian-distributed, the LMMSE estimation error is not necessarily independent of the channel estimate. 
We therefore adopt the UatF bound \cite{massive_mimo_foundations}, which is applicable to any channel estimator. Under this bound, the component of the received signal corresponding to the mean effective channel is treated as the desired signal, while the remaining terms are interpreted as interference and noise. The resulting
signal-to-interference-plus-noise rartio (SINR) of UE $k$ is
\begin{equation}
\overline{\mathrm{SINR}}_{k}
=
\frac{
(\alpha_2\alpha_1)^2\rho_k|\mu_{k}|^2
}{
(\alpha_2\alpha_1)^2
\displaystyle\sum_{i=1}^{K}\rho_i\zeta_{k,i}
+
\chi_{k}
-
(\alpha_2\alpha_1)^2\rho_k|\mu_{k}|^2
},
\label{eq:uatf_sinr}
\end{equation}
where
\begin{align}
\mu_{k}
&\triangleq
\E\left\{
\mathbf v_{k,l}^{\mathrm H}\hb_{k,l}
\right\},
\label{eq:mu_def}\\
\zeta_{k,i}
&\triangleq
\E\left\{
\left|
\mathbf v_{k,l}^{\mathrm H}\hb_{i,l}
\right|^2
\right\},
\label{eq:zeta_def}\\
\chi_{k}
&\triangleq
\E\left\{
\mathbf v_{k,l}^{\mathrm H}
\mathbf C_{\tilde{\mathbf n}\tilde{\mathbf n}}
\mathbf v_{k,l}
\right\}.
\label{eq:chi_def}
\end{align}
The expectations in \eqref{eq:mu_def}--\eqref{eq:chi_def} are taken over
the channel realizations and the corresponding pilot observations in any active CB. $\mathbf C_{\tilde{\mathbf n}\tilde{\mathbf n}}$ in \eqref{eq:chi_def} is the conditional covariance of the effective noise in \eqref{eq:effective_noise}, which is given by
\begin{equation}
\begin{aligned}
\mathbf C_{\tilde{\mathbf n}\tilde{\mathbf n}}
&\triangleq
\E\left\{
\tilde{\mathbf n}[m]
\tilde{\mathbf n}[m]^{\mathrm H}
\mid\mathcal H
\right\} \\
&=
(\alpha_2\alpha_1)^2\sigma^2\mathbf I_N
+
\alpha_2^2\mathbf D_{\bar\eta_1\mid\mathcal H}
+
\mathbf D_{\bar\eta_2\mid\mathcal H}.
\end{aligned}
\label{eq:Ceff_data}
\end{equation}
The achievable SE of UE $k$ is
\begin{equation}
\overline{\mathrm{SE}}_{k}
=
\beta\,
\frac{\tau_u}{\tau_c}
\log_2\left(
1+\overline{\mathrm{SINR}}_{k}
\right),
\label{eq:se}
\end{equation}
where $\tau_c=\nu_{\mathrm{CB}}\lambda_{\mathrm{CB}}$ is the number of
resource elements in one CB,
$\tau_u=\tau_c-\tau_p=(\nu_{\mathrm{CB}}-1)\lambda_{\mathrm{CB}}$ is the
number of resource elements used for uplink data transmission. The sum SE is 
\begin{equation}
\overline{\mathrm{SE}}=\sum_{k=1}^K\overline{\mathrm{SE}}_{k}.
\label{eq:sum_se}
\end{equation}

\section{Numerical Results}
\label{sec:numerical_results}

\par In this section, simulation results are presented to demonstrate the impact of double quantization. To this end, we aim to answer the following questions:
\begin{itemize}
    \item How does the resolution at each quantization stage affect the sum SE?
    \item How do pilot contamination and the number of UEs affect the sum SE under double quantization?
    \item How does the sum SE vary across different combining schemes?
\end{itemize}

We consider NMR, RZF, and MMSE combining as given in \cite{massive_mimo_foundations}, using the LMMSE channel estimates derived in Section~\ref{sec:est}. For NMR combining, we use
\begin{equation}
\mathbf v_{k,l}^{\mathrm{NMR}}
=
\frac{\Hb_{k,l}}{\|\Hb_{k,l}\|^2},
\label{eq:mr_combiner}
\end{equation}
which provides a tighter UatF bound than the classical MR combiner \cite{massive_mimo_foundations}. For RZF combining, $\mathbf v_{k,l}^{\mathrm{RZF}}$ is
the $k$th column of
\begin{equation}
\mathbf V_l^{\mathrm{RZF}}
=
\left(
\hat{\bar{\mathbf H}}_l
\mathbf P
\hat{\bar{\mathbf H}}_l^{\mathrm H}
+
\sigma^2\mathbf I_N
\right)^{-1}
\hat{\bar{\mathbf H}}_l\mathbf P,
\label{eq:rzf_combiner}
\end{equation}
where
$\hat{\bar{\mathbf H}}_l
\triangleq
[\Hb_{1,l},\dots,\Hb_{K,l}]$ and
$\mathbf P\triangleq\diag(\rho_1,\dots,\rho_K)$.
Similarly, the MMSE combiner $\mathbf v_{k,l}^{\mathrm{MMSE}}$ is the
$k$th column of
\begin{equation}
\mathbf V_l^{\mathrm{MMSE}}
=
\left(
\hat{\bar{\mathbf H}}_l
\mathbf P
\hat{\bar{\mathbf H}}_l^{\mathrm H}
+
\mathbf C_{\mathrm{err}}
+
\sigma^2\mathbf I_N
\right)^{-1}
\hat{\bar{\mathbf H}}_l\mathbf P,
\label{eq:mmse_combiner}
\end{equation}
where
$\mathbf C_{\mathrm{err}}
\triangleq
\sum_{i=1}^{K}\rho_i\mathbf C_{\mathrm{err},i}$
is the total estimation-error covariance, with $\mathbf C_{\mathrm{err},i}$
given in \eqref{eq:errcov}. For each combining scheme, the achievable
SE of UE $k$ is obtained by substituting the corresponding combining
vector into \eqref{eq:uatf_sinr} and then using \eqref{eq:se}. The expectations in
\eqref{eq:mu_def}--\eqref{eq:chi_def} are evaluated numerically by Monte Carlo averaging. 
\par We consider a massive MIMO system in which a single BS equipped with $N=100$ antennas serves $K$ single-antenna UEs. The UEs are uniformly distributed over a $\SI{250}{\meter}\times\SI{250}{\meter}$ square area, subject to a minimum distance of \SI{35}{\meter} from the BS. The large-scale channel gain of UE $i$ is modeled as $-35.3-37.6\log_{10}(d_i)+z_i$~\si{\decibel}, where $d_i$ denotes the
distance from UE $i$ to the BS in meters, and $z_i\sim\mathcal N(0,10^2)$ accounts for log-normal shadow fading. The spatial correlation matrix $\mathbf R_i$ is generated according to the local scattering model in \cite{massive_mimo_foundations}, using a Gaussian angular distribution with an angular standard deviation of \SI{10}{\degree}. The OFDM waveform consists of $M=2048$ subcarriers with a subcarrier spacing of $\Delta f=\SI{30}{\kilo\hertz}$. We assume each CB spans $\nu_{\mathrm{CB}}=20$ OFDM symbols in all our results, and we assume the first OFDM symbol in each CB is occupied by pilot symbols, i.e., $\tau_p=\lambda_{\mathrm{CB}}$.
The number of active subcarriers is fixed to $L_{\mathrm{act}}\lambda_{\mathrm{CB}}=1600$ in all scenarios, corresponding to a spectral occupancy ratio of $\beta=25/32$. 
This is a practically relevant choice since a \SI{50}{\mega\hertz} channel bandwidth with a \SI{30}{\kilo\hertz} subcarrier spacing corresponds to the NR FR1 transmission configuration\ with $1596$ subcarriers (excluding null subcarriers) \cite{nr_active_subcarriers}.
All UEs transmit with $\rho_i=\SI{100}{\milli\watt}$, and the noise variance $\sigma^2$ is computed over the sampling bandwidth $M\Delta f$ with a receiver noise figure of \SI{7}{\decibel}.

\begin{figure}[t]
\centering
\includegraphics[width=\columnwidth]{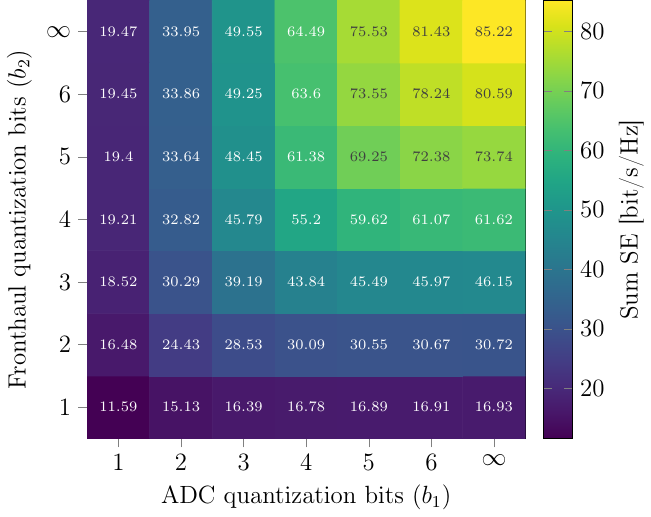}
\caption{Sum SE of MMSE combining versus the ADC and fronthaul resolutions $(b_1,b_2)$ for $K=20$ and $\lambda_{\mathrm{CB}}=32$, where $b=\infty$ corresponds to unquantized processing.}
\label{fig:heatmap} 
\vspace{-6mm}
\end{figure}
\par In Fig.~\ref{fig:heatmap}, we report the sum SE in \eqref{eq:sum_se} achieved with MMSE combining for different ADC and fronthaul quantization resolutions, with $K=20$ and $\lambda_{\mathrm{CB}}=\tau_p=32$. As expected, increasing the resolution of either quantization stage increases the sum SE. More interestingly, we observe that increasing the resolution of only one stage provides a limited gain when the other stage operates at a low resolution. For example, under two-bit fronthaul quantization, i.e., $b_2=2$, increasing the ADC resolution from two to six bits improves the sum SE by about \SI{6}{bit/s/Hz}. In contrast, increasing both quantization resolutions to six bits improves the sum SE by about \SI{54}{bit/s/Hz}. This is because the quantization stage with the lower resolution distorts the signal significantly, which limits the performance regardless of the other quantization resolution.
These results also show that the fronthaul and ADC quantization resolutions have a similar impact on the sum SE. For example, the sum SE difference between $(b_1,b_2)=(4,5)$ and $(b_1,b_2)=(5,4)$ is only about \SI{2}{bit/s/Hz}. Moreover, when both quantization stages use six-bit resolution, more than 90\% of the SE relative to the unquantized case is achieved. This shows that six-bit quantization at both stages provides a good trade-off between SE and quantization resolution; utilizing only a part of fronthaul bandwidth.

\begin{figure}[t]
\centering
\includegraphics[
    width=0.95\columnwidth,
    keepaspectratio
]{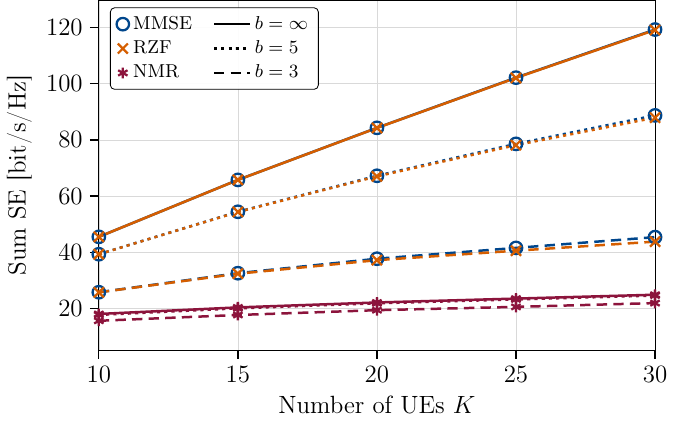}
\caption{Sum SE of NMR, RZF, and MMSE combining versus the
number of UEs $K$ for $b_1=b_2=b\in\{3,5,\infty\}$, with
$\lambda_{\mathrm{CB}}=\tau_p=32$ and without pilot contamination.}
\label{fig:users_nopc}
\vspace{-4mm}
\end{figure}
\begin{figure}
\centering
\includegraphics[
    width=0.95\columnwidth,
    keepaspectratio
]{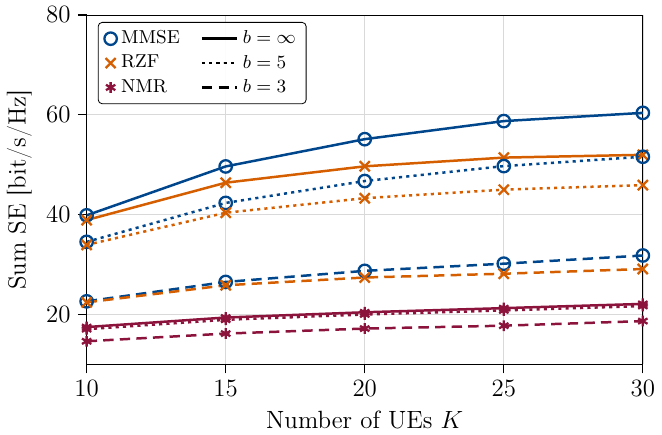}
\caption{Sum SE of NMR, RZF, and MMSE combining versus the
number of UEs $K$ for $b_1=b_2=b\in\{3,5,\infty\}$, with
$\lambda_{\mathrm{CB}}=\tau_p=8$ and pilot contamination.}
\label{fig:users_pc}
\vspace{-6mm}
\end{figure}

\par In Fig.~\ref{fig:users_nopc} and Fig.~\ref{fig:users_pc}, we compare the sum SE as a function of the number of UEs for different combining methods and quantization resolutions. In Fig.~\ref{fig:users_nopc}, we have $\lambda_{\mathrm{CB}}=32$ and $\tau_p=\lambda_{\mathrm{CB}}$. Since $K\leq\tau_p$ for all considered values of $K$, orthogonal pilot vectors can be assigned to every UE. Thus, in \eqref{eq:pilot_observation}, $i$ can be replaced by the user index $k$. In this setup, the channel estimates are accurate, which limits the additional gain provided by MMSE over RZF. Since $N \gg K$, RZF and MMSE effectively mitigate the interference terms in \eqref{eq:zeta_def}. Their sum SE therefore increases almost linearly with $K$ and remains higher than that of NMR. The quantization distortion then becomes the dominant impairment in \eqref{eq:uatf_sinr}, which makes their SE more sensitive to the quantization resolution.

In Fig.~\ref{fig:users_pc}, we plot the sum SE as a function of the number of UEs for $\lambda_{\mathrm{CB}}=\tau_p=8$. When $K>\tau_p$, pilot reuse is unavoidable and creates pilot contamination. As $K$ increases, more UEs share the available pilot vectors, which degrades the accuracy of the channel estimates. Due to inter-user interference and inaccurate channel estimates, the sum SE increases only marginally as $K$ increases. We also observe that the gap between MMSE and RZF becomes visible. As expected, NMR has the lowest sum SE. 

\section{Conclusion}
In this paper, we investigated the achievable uplink SE of a massive MIMO-OFDM system where the received signal is quantized first by the low-resolution ADCs in the time domain and then by the fronthaul quantizer. We derived an LMMSE channel estimator from pilot signals affected by both quantization stages and obtained an achievable SE using the UatF bound. The numerical results show that the two quantization stages have a comparable impact on the sum SE. When the quantization resolutions differ, the lower resolution limits the sum SE significantly. It is therefore recommended to use similar resolutions in both stages. In our numerical experiments, MMSE and RZF combining achieve a similar sum SE when orthogonal pilots are assigned to all UEs, whereas MMSE outperforms RZF under pilot contamination since the channel estimation error is taken into account in the combiner design. In our future work, we aim to introduce quantization-aware receivers. We also plan to expand our analysis to cell-free massive MIMO systems.
\bibliographystyle{IEEEtran}
\bibliography{bibtex/bib/references}

@article{massive_mimo_foundations,
  author  = {E. Bj{\"o}rnson and J. Hoydis and L. Sanguinetti},
  title   = {{Massive MIMO} Networks: Spectral, Energy, and Hardware Efficiency},
  journal = {Found. Trends Signal Process.},
  volume  = {11},
  number  = {3--4},
  pages   = {154--655},
  year    = {2017}
}

@article{aqnm_massive_mimo_rate,
  author  = {L. Fan and S. Jin and C.-K. Wen and H. Zhang},
  title   = {Uplink Achievable Rate for {Massive MIMO} Systems With Low-Resolution {ADC}},
  journal = {IEEE Commun. Lett.},
  volume  = {19},
  number  = {12},
  pages   = {2186--2189},
  month   = dec,
  year    = {2015}
}

@article{aqnm_rician_se,
  author  = {J. Zhang and L. Dai and S. Sun and Z. Wang},
  title   = {On the Spectral Efficiency of {Massive MIMO} Systems With Low-Resolution {ADCs}},
  journal = {IEEE Commun. Lett.},
  volume  = {20},
  number  = {5},
  pages   = {842--845},
  month   = may,
  year    = {2016}
}

@article{wideband_quantized_mimo_ofdm,
  author  = {C. Studer and G. Durisi},
  title   = {Quantized {Massive MU-MIMO-OFDM} Uplink},
  journal = {IEEE Trans. Commun.},
  volume  = {64},
  number  = {6},
  pages   = {2387--2399},
  month   = jun,
  year    = {2016}
}

@inproceedings{oversampled_quantized_mimo_ofdm,
  author    = {M. Ma and N. T. Nguyen and I. Atzeni and M. Juntti},
  title     = {Analysis of Oversampling in Uplink {Massive MIMO-OFDM} With Low-Resolution {ADCs}},
  booktitle = {Proc. IEEE 24th Int. Workshop Signal Process. Adv. Wireless Commun. (SPAWC)},
  pages     = {626--630},
  month = sep,
  year      = {2023}
}

@inproceedings{wideband_cell_free_adc,
  author    = {Demir, {\"O}. T. and Elbir, A. M. and Bj{\"o}rnson, E.},
  title     = {Cell-Free {Massive MIMO-OFDM} With Low-Resolution {ADCs}},
  booktitle = {Proc. 23rd Int. Symp. Model. Optim. Mobile, Ad Hoc, Wireless Netw. (WiOpt)},
  pages     = {1--7},
  month = may,
  year      = {2025}
}

@article{fronthaul_quantization,
  author  = {G. Femenias and F. Riera-Palou},
  title   = {Fronthaul-Constrained Cell-Free {Massive MIMO} With Low Resolution {ADCs}},
  journal = {IEEE Access},
  volume  = {8},
  month = jun,
  pages   = {116195--116215},
  year    = {2020}
}

@inproceedings{double_quantization_ofdm,
  author    = {Demir, {\"O}. T. and Bj{\"o}rnson, E.},
  title     = {Joint Impact of {ADC} and Fronthaul Quantization in Cell-Free {Massive MIMO-OFDM} Uplink},
  booktitle = {Proc. Asilomar Conf. Signals, Syst., Comput.},
  pages     = {577--581},
  month = oct,
  year      = {2025}
}

@article{coherence_block_ofdm,
  author  = {Y. Wu and L. Sanguinetti and M. F. Keskin and U. Gustavsson and {Graell i Amat}, A. and H. Wymeersch},
  title   = {Uplink Cell-Free {Massive MIMO OFDM} With Phase Noise-Aware Channel Estimation: Separate and Shared Local Oscillators},
  journal = {IEEE Trans. Wireless Commun.},
  volume  = {25},
  pages   = {1656--1673},
  year    = {2026}
}

@article{bussgang_aqnm,
  author  = {Demir, {\"O}. T. and Bj{\"o}rnson, E.},
  title   = {The {Bussgang} Decomposition of Nonlinear Systems: Basic Theory and {MIMO} Extensions [Lecture Notes]},
  journal = {IEEE Signal Process. Mag.},
  volume  = {38},
  number  = {1},
  pages   = {131--136},
  month   = jan,
  year    = {2021}
}

@techreport{nr_active_subcarriers,
  author      = {{3GPP}},
  title       = {{NR}; Base Station ({BS}) Radio Transmission and Reception},
  institution = {{3GPP}},
  type        = {Technical Specification},
  number      = {TS 38.104},
  month       = may,
  year        = {2024},
  note        = {Version 17.13.0, Release 17}
}

@article{towards_6g_mimo,
  author  = {E. Bj{\"o}rnson and C.-B. Chae and R. W. Heath, Jr. and
             T. L. Marzetta and A. Mezghani and L. Sanguinetti and
             F. Rusek and M. R. Castellanos and D. Jun and
             {\"O}. T. Demir},
  title   = {Towards {6G} {MIMO}: Massive spatial multiplexing, dense
             arrays, and interplay between electromagnetics and processing},
  journal = {arXiv preprint arXiv:2401.02844},
  month = jan,
  year    = {2024},
}

@article{cell_free_uniform_quantization,
  author  = {M. Bashar and K. Cumanan and A. G. Burr and H. Q. Ngo and M. Debbah and P. Xiao},
  title   = {Max--Min Rate of {Cell-Free Massive MIMO} Uplink With Optimal Uniform Quantization},
  journal = {IEEE Trans. Commun.},
  volume  = {67},
  number  = {10},
  pages   = {6796--6815},
  month   = oct,
  year    = {2019}
}

@article{cell_free_mmwave_adc_fronthaul,
  author  = {I.-S. Kim and M. Bennis and J. Choi},
  title   = {{Cell-Free} mmWave Massive MIMO Systems With Low-Capacity Fronthaul Links and Low-Resolution {ADC/DACs}},
  journal = {IEEE Trans. Veh. Technol.},
  volume  = {71},
  number  = {10},
  pages   = {10512--10526},
  month   = oct,
  year    = {2022}
}

@article{fronthaul_rate_constraint_quantized_mimo,
  author  = {Y. Ettefagh and S. R. Aghdam and G. Durisi and S. Jacobsson and M. Coldrey and C. Studer},
  title   = {Performance of Quantized Massive MIMO With Fronthaul Rate Constraint Over Quasi-Static Channels},
  journal = {IEEE Access},
  volume  = {11},
  pages   = {56935--56950},
  month   = jun,
  year    = {2023}
}

\end{document}